\documentclass{PoS}

\def\mpi2{m_\pi^2}
\def\mK2{m_K^2}

\newcommand{\bea}{\begin{eqnarray}}
\newcommand{\eea}{\end{eqnarray}}
\newcommand{\be}{\begin{equation}}
\newcommand{\ee}{\end{equation}}

\def\maths#1{$#1$}

\newcommand{\Tr}{\mbox{Tr}}

\def\lsim{\raise0.3ex\hbox{$<$\kern-0.75em\raise-1.1ex\hbox{$\sim$}}}
\newsavebox{\DERIVBOXZLM}
\savebox{\DERIVBOXZLM}[2.5em]{$\Longrightarrow\hspace{-1.5em}
\raisebox{.2ex}{*}
\hspace{-.7em}\raisebox{-.8ex}{\scriptsize lm}\hspace{.7em}$}

\newcommand{\psibar}{\overline{\psi}}

\newcommand{\pbp}{\langle\psibar\psi\rangle}

\usepackage{epsf}
\title{QCD at nonzero temperature and density}

\ShortTitle{QCD at nonzero temperature and density}

\author{\speaker{Ludmila Levkova}\\
        Physics Department, University of Utah, Salt Lake City, UT 84112, USA\\
        E-mail: \email{ludmila@physics.utah.edu}}
\abstract{The properties of hot hadronic matter are of great importance to the studies of heavy-ion collisions, cosmology and compact star formation. I briefly outline the current methods in use in the lattice simulations of QCD thermodynamics at zero and nonzero density. I discuss the most recent results for the QCD phase transition, critical behavior and the equation of state.}

\FullConference{The XXIX International Symposium on Lattice Field Theory\\
                 July 11-16, 2011\\
                 Squaw Valley, Lake Tahoe, California}

\begin{document}
\section{Introduction}

At low temperatures and densities the quarks and the gluons are "confined"
in colorless states such as the nucleons and the mesons. In other words, there are no free
quarks or gluons detected experimentally. The question of what happens to ordinary 
hadronic matter at extreme conditions (very high temperatures or densities) has been 
recently answered by the heavy-ion collision (HIC) experiments at RHIC and LHC. At these experiments,
heavy nuclei accelerated close to the speed of light collide to form a new state of matter 
called the quark-gluon plasma (QGP).    
Prior to its discovery, the existence of the QGP had been speculated about for more than 30 years \cite{qgp}
on the basis of the expectation of asymptotic freedom of QCD \cite{asym} at high energies. 
The properties of the QGP 
are important for the study of a variety of physical phenomena. For example, up to several microseconds 
after the big bang all the hadronic matter in the Universe was in a form of QGP. Hence, the equation of
state (EOS) of QGP plays a role in the expansion of the early Universe and  may influence
the subsequent structure formation. The QGP also may exist in the cores of neutron stars (at very high
baryon density) and may even form quark stars. And lastly, for HIC experiments 
the EOS of QGP is an essential 
input to the hydrodynamic equations used to simulate the expansion of the created plasma.
The initial temperatures occurring  at the HIC 
are not more than five times
higher than the transition temperature $T_c$ between ordinary matter and quark-gluon plasma. 
In this temperature range the quark-gluon plasma is in a nonperturbative regime and only a 
nonperturbative tool, such as lattice QCD, seems to provide an adequate description.

Lattice QCD is simply the discretized version of QCD on a space-time lattice of dimensions
$N_s^3\times N_t$. The temperature on the lattice is defined as $T=1/(aN_t)$, where $a$ is 
the lattice spacing. To change the temperature on the lattice one can fix $N_t$ and vary $a$.
This type of temperature variation preferably should occur on lines of constant physics (LCP). A different
strategy would be to fix the lattice spacing and change $T$ by varying $N_t$. This way staying on 
an LCP is automatically guaranteed. The continuum limit  at a particular $T$ is taken
as $N_t\rightarrow\infty$ (i.e., $a\rightarrow0$). Generally, to insure that a lattice calculation
will not suffer from large finite volume effects, the following inequality should be 
fulfilled:
\be
a\ll M_h^{-1}\ll aN_s,
\ee
where $M_h$ is the mass of some characteristic light hadron. The temperature on the lattice can be  
considered high when $TM_h^{-1}\sim O(1)$, which means that the above inequality in this case 
is modified to:
\be
1/N_t\ll1\ll N_s/N_t.
\ee
Current simulations are done with $N_t=6-16$ and aspect ratios $N_s/N_t=3-4$. The discretization effects
at finite $a$ will strongly depend on the choice of lattice action. There are two main types of 
fermion lattice actions in use: staggered types \cite{ks} and Wilson types \cite{wil} (including chiral formulations \cite{dwf}), 
with the former being the choice with
the largest investment of computing resources thus far at high $T$.

\section{Staggered actions and discretization effects}

The most advanced nonzero temperature calculations so far have been done with staggered fermions.
This type of fermion discretization is one of the most computationally affordable and thus it still 
remains as the most popular choice for the expensive high-temperature
calculations. Another advantage of 
the staggered fermions is that they retain a $U(1)\times U(1)$ symmetry (for one staggered flavor), 
which is a remnant of the full chiral symmetry
group. This remaining symmetry protects the chiral limit from an additive renormalization
in the quark mass. The main drawback of the staggered type of discretization is that for
a single staggered flavor there are 4 "tastes" of Dirac fermions appearing in the continuum limit. To
deal with this taste proliferation, the usual strategy is to work with the fourth root of the fermion 
determinant instead of the full one in the Monte Carlo generation of gauge ensembles. The hope is that
this procedure effectively reduces the number of tastes to one per flavor, although this has not been
proven rigorously. Still, there is a growing amount of numerical \cite{rootn} and theoretical \cite{roott}
evidence that
the fourth root procedure gives the correct number of flavors in the continuum limit, provided it is taken
before the chiral limit. At finite lattice spacing the taste symmetry is broken, which means that
the staggered hadron multiplets are nondegenerate. For example, there are 16 staggered pions
and the fact that they have nondegenerate masses can distort the physics at not-too-high temperatures
below the crossover,
where they are expected to dominate it.

\begin{figure}[t]
\begin{tabular}{cc}
 \epsfxsize=72mm
  \epsfbox{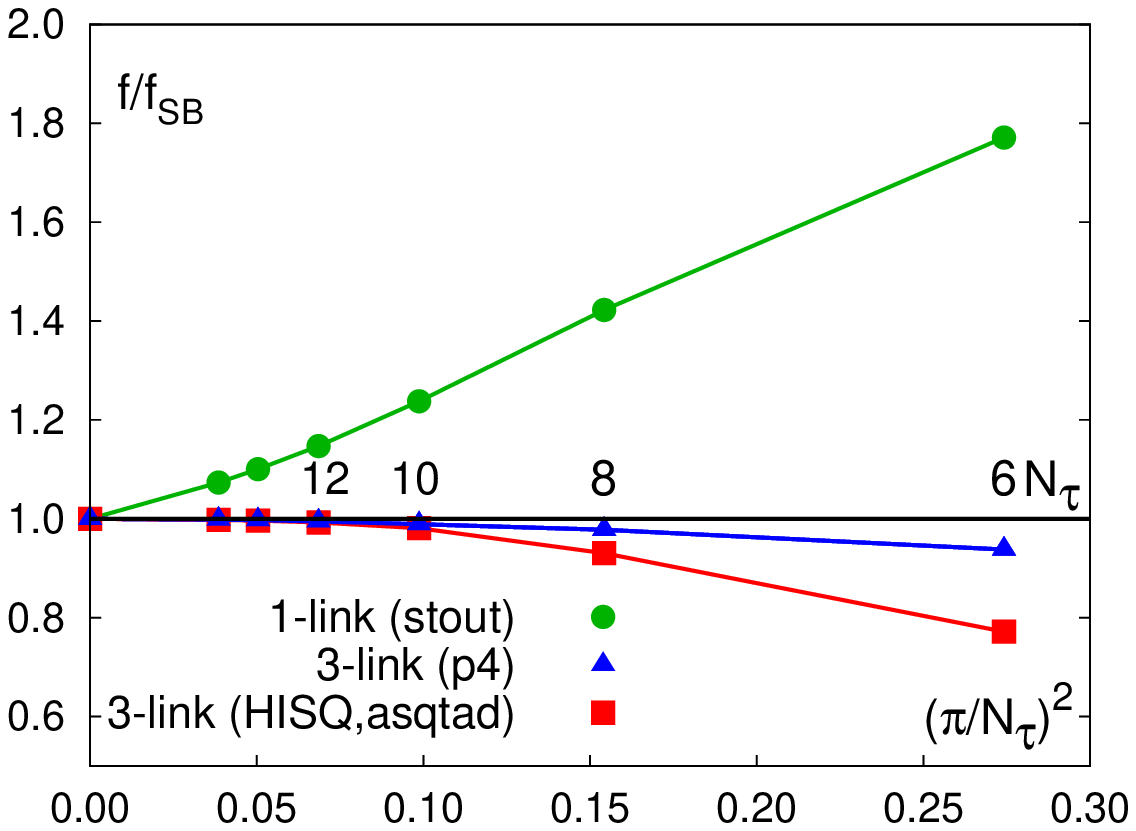}
&
  \epsfxsize=72mm
  \epsfbox{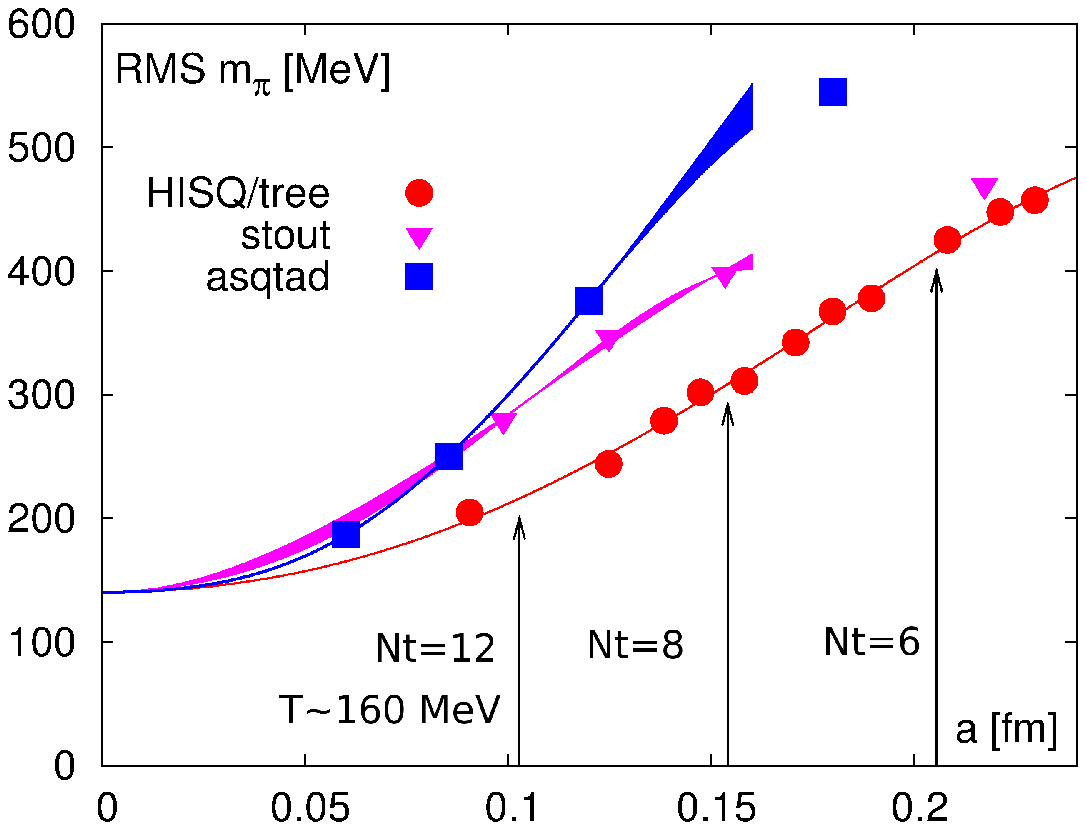}
\end{tabular}
\caption{(Left) The free energy normalized to the Stefan-Boltzmann value {\it vs.} the square of the inverse of temporal extent in
the free theory case. (Right) The square of the RMS pion mass {\it vs.} the lattice spacing \cite{abazavov}. It is assumed that the Goldstone pion
mass is tuned to its physical value. 
The vertical lines denote the approximate lattice spacings 
at which the transition happens for different $N_t$'s.}
\label{fig:stagg}
\end{figure}

There are a number of improved staggered actions in use: the $p4$ \cite{p4}, asqtad \cite{asq}, stout 
\cite{stout} and the HISQ \cite{hisq} action,
which deal with the discretization effects in somewhat different ways. The $p4$ action has an improved 
quark dispersion relation to $O(p^4)$, where $p$ is the quark momentum. The asqtad action also has this property but
furthermore, it has all tree-level lattice artifacts to $O(a^2)$ removed, and its coefficients are tadpole-improved.
The stout action has the taste symmetry violation reduced through unitary link smearing. And finally, the HISQ
action combines all of the above improvements. 
Comparisons of the sizes of the lattice artifacts among the above actions are shown in Fig.~\ref{fig:stagg}.
In the left panel the approach to the continuum limit of the free energy in the free fermion case is presented. 
The stout action has a slower approach to the Stefan-Boltzmann continuum value than the rest. The closer the value of the free energy
from a particular action is
to the continuum the smaller are the UV cutoff effects for this action in the high-temperature regime above the phase transition.
However, in the regime of low temperature or around the transition, the taste symmetry violation would seem to play 
a more important role in determining the discretization effects. If we choose the root-mean-square (RMS) pion mass as a measure
of the taste symmetry violation defined as
\be
m^{RMS}_\pi=\sqrt{
  \frac{1}{16}\left(m_{\gamma_5}^2+m_{\gamma_0\gamma_5}^2
  +3m_{\gamma_i\gamma_5}^2+3m_{\gamma_i\gamma_j}^2
  +3m_{\gamma_i\gamma_0}^2+3m_{\gamma_i}^2
  +m_{\gamma_0}^2+m_{1}^2\right)}\,,
\ee
where the sum is over the squares of the masses of the members of the staggered pion multiplet,
then the HISQ action
appears to have the smallest taste symmetry violations of all at any $a$'s. On the other hand, the stout action  
outperforms the asqtad action for coarser lattices.    
 
\section{QCD at zero density}
\begin{figure}[t]
\begin{tabular}{cc}
 \epsfxsize=72mm
  \epsfbox{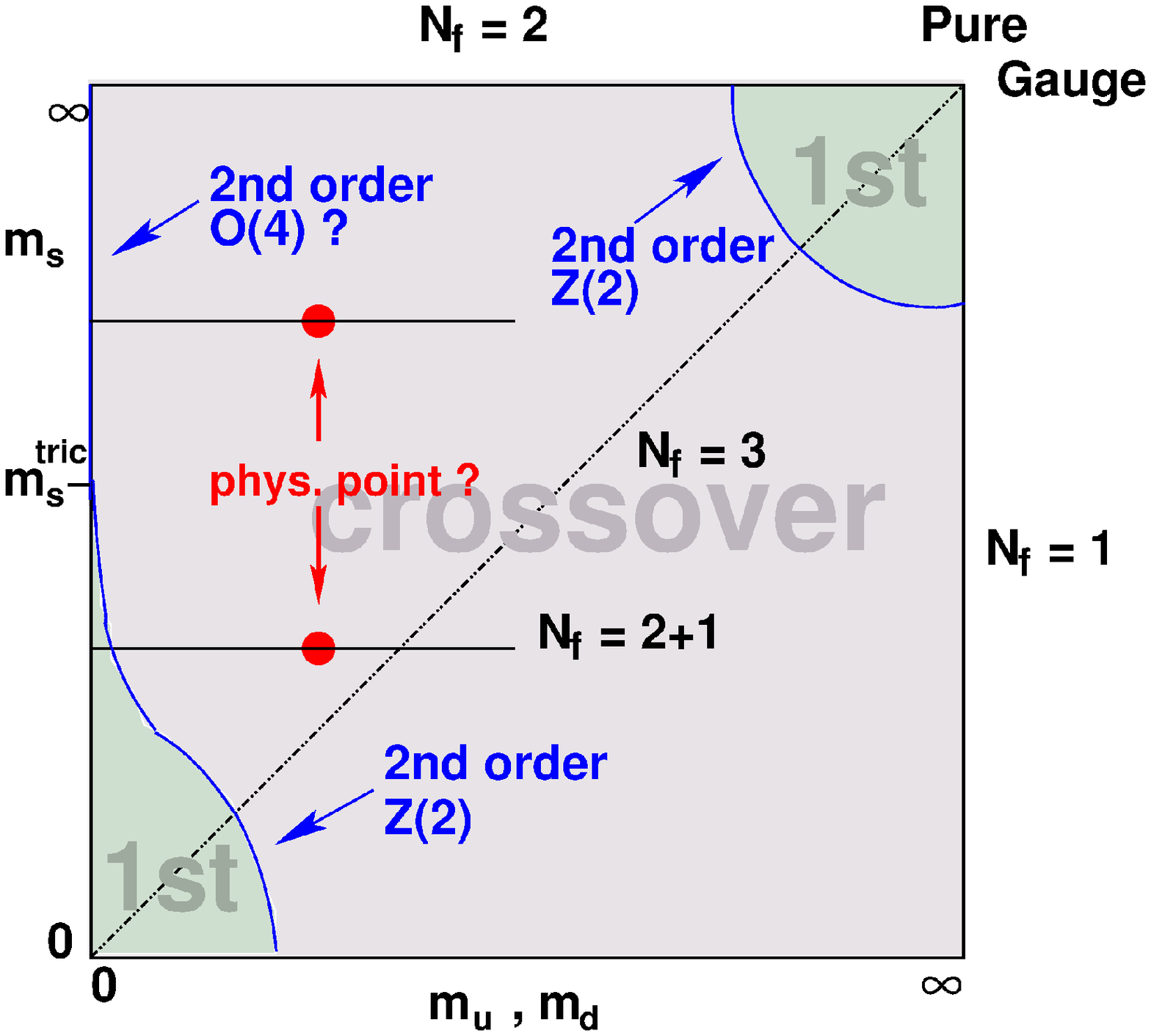}
&
  \epsfxsize=72mm
  \epsfbox{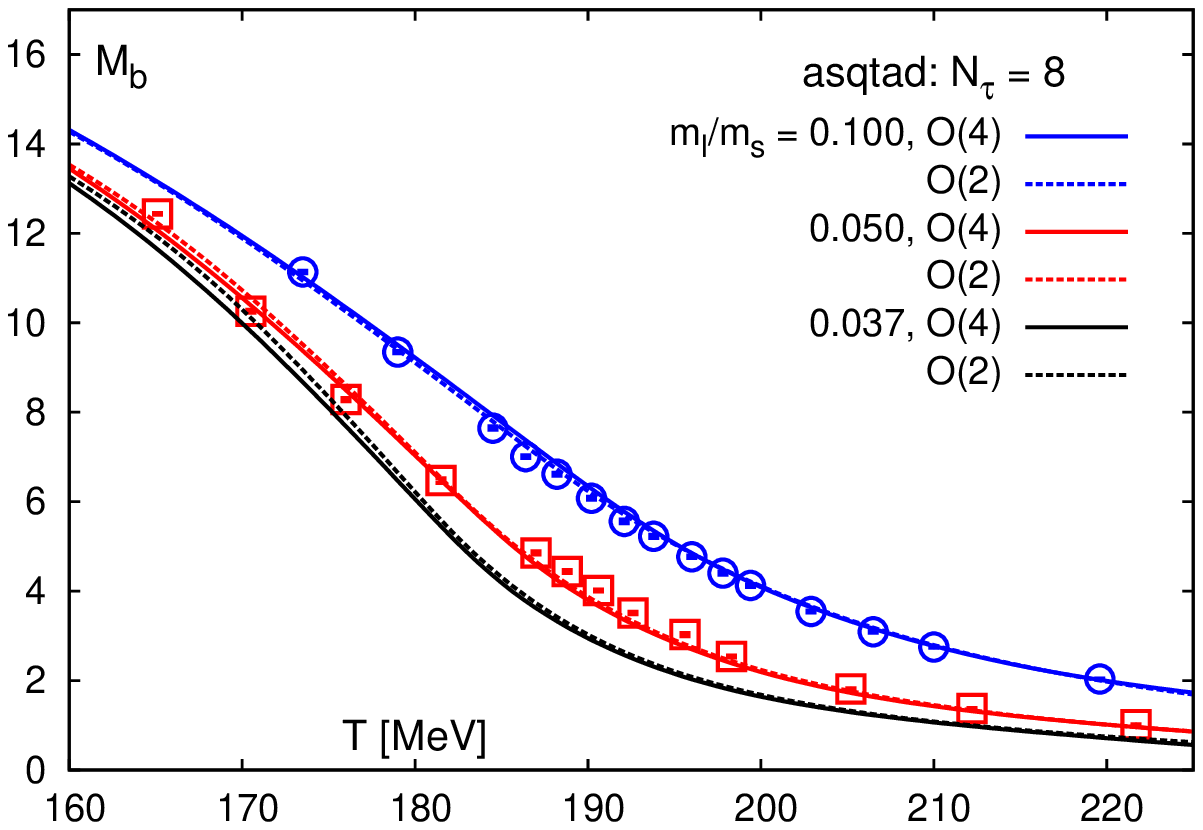}
\end{tabular}
\caption{(Left) The QCD phase diagram at zero chemical potential. (Right) Fits with $O(N)$ scaling magnetic equations to the chiral
order parameter \cite{abazavov}. }
\label{fig:diag}
\end{figure}

The widely accepted conjecture for the QCD phase diagram in the plane of the strange quark mass $m_s$ {\it vs.} the light quark masses
$m_{u,d}$ is shown in Fig~\ref{fig:diag} (left panel).
At a range of low or high quark masses 
(shown as green areas in the lower-left and upper-right corners), the expectation is that
the phase transition at high $T$ is of 1st order. In the first case it should have the character of chiral 
symmetry restoration and in the second --- of a deconfining type. Both of these areas are surrounded by
critical lines of 2nd order transition. At zero $m_{u,d}$ and sufficiently large $m_s$, the 2nd order critical line
changes from $Z(2)$-type to $O(4)$-type at the tricritical point $m_s^{\rm tric}$. The current consensus is that 
the physical point is somewhere in the vast area of the crossover region \cite{cross}. It is important to study the phase diagram quantitatively,
especially when locating the physical point in relation to the low-mass region of 1st order transition.
If the physical point lies low, when taking the chiral limit one encounters a $Z(2)$ 2nd order phase transition at nonzero mass.
On the other hand
when it lies higher, the 2nd order transition is expected to be of $O(4)$-type. Establishing which particular behavior 
is true determines the type of scaling functions one should use to describe data at low quark masses. In a recent study \cite{mcrit} with
2+1 flavors of stout fermions, 
it was found that the 2nd order critical behavior starts at quark masses $\lsim12\%$ of the physical quark masses.
This 
suggests that the physical point is relatively far from the region of the 1st order transition and 
the chiral phase transition is probably of $O(4)$-type. This year,
a new study \cite{htding} with 3 light degenerate flavors 
of HISQ fermions reached 
similar conclusions --- the critical quark masses are found to be $\lsim10\%$ of the 
physical light quark mass. Another new study \cite{dsmith} with 3 degenerate flavors confirms the $Z(2)$-type of scaling behavior as expected from the
qualitative picture of the phase diagram.

The upper-right corner of the phase diagram has also been explored numerically. The WHOT-QCD collaboration has done a recent study \cite{heavy}
of the heavy-quark region starting from pure $SU(3)$ configurations (i.e., infinite quark masses) and using reweighting combined with
hopping parameter expansion to reduce the quark masses and map the critical line. This year they extended this work by adding
nonzero chemical potential \cite{hsaito}.
       
\subsection{QCD and universal scaling}
\begin{figure}[t]
\begin{tabular}{cc}
 \epsfxsize=72mm
  \epsfbox{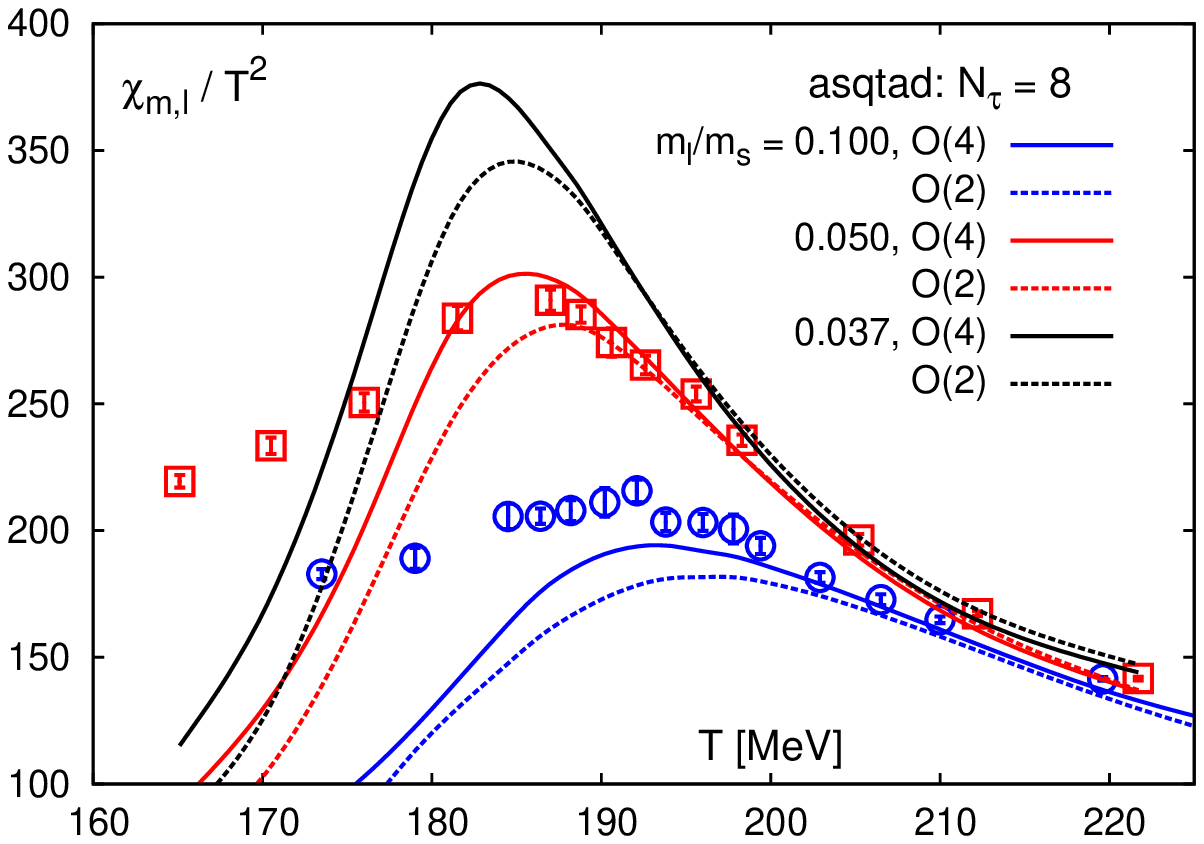}
&
  \epsfxsize=72mm
  \epsfbox{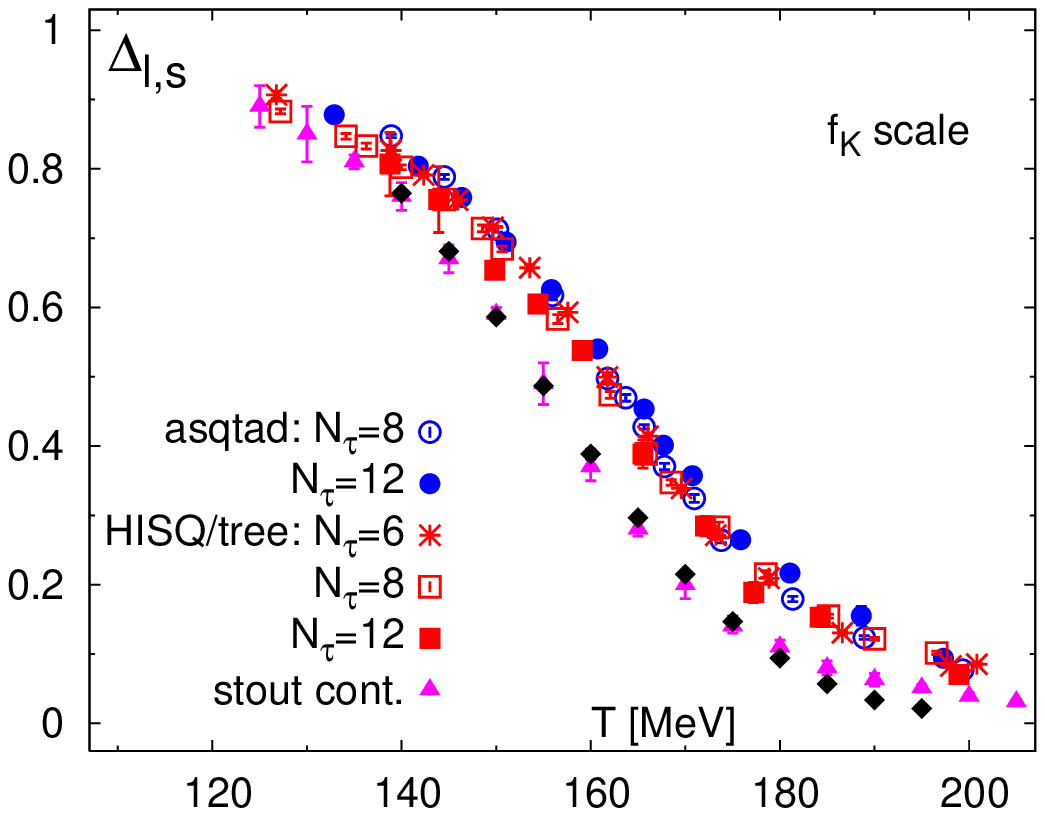}
\end{tabular}
\caption{(Left) Comparison of asqtad data for $\chi_{m,l}$ with the differentiated scaling O(N) curves \cite{abazavov}.  (Right) Comparison
between the HotQCD and WB data for the subtracted chiral condensate \cite{abazavov}. The continuum and physical quark limit for 
the WB data is denoted by pink triangles. 
The same for the HotQCD data is denoted with black diamonds.}
\label{fig:scaling}
\end{figure}
As described in the previous subsection, there is an increasing amount of evidence that in the chiral limit of 2+1
flavor QCD the phase transition is of 2nd order in the $O(4)$ universality class. Close to the chiral transition (in temperature
and quark masses) the free energy can be written as a sum of a singular part $f_s$ and a regular part $f_r$:
\be
\frac{f}{T^4}(T,m_l,m_s) = h^{1+1/\delta}f_s(z) + f_r(T,H,m_s),
\ee   
where $z=t/h^{1/(\beta\delta)}$ is a scaling variable and $\beta$ and $\delta$ are critical exponents of the $O(N)$
universality class. The reduced temperature $t$ and the symmetry breaking field $h$ are defined below: 
\be
t = \frac{1}{t_0} \left( \frac{T-T_c^0}{T_c^0} +
\sum_{q=l,s}\kappa_q\left(\frac{\mu_q}{T}\right)^2 +
\kappa_{ls} \frac{\mu_l}{T}\frac{\mu_s}{T} \right)
 \quad , \quad
h= \frac{1}{h_0} H \quad , \quad
H= \frac{m_l}{m_s},
\ee
where in the general case $t$ depends on the light and heavy quark chemical potentials $\mu_l$ and $\mu_s$ (in the 2+1 flavor case)
and $T_c^0$ is the true chiral critical temperature. 
The dynamics around the phase transition is governed by the scale invariant singular part $f_s$, which means that
to extract the critical temperature it is best to choose observables sensitive to it (the regular part $f_r$
gives rise to scaling violations in these observables). Such an observable
is of course the chiral order parameter, which has the critical behavior $\langle \bar \psi \psi \rangle_l \sim  m_{l}^{1/\delta}$
at fixed $z$. Another observable sensitive to $f_s$ is the chiral susceptibility 
$\chi_{m,l} =  \partial\langle\bar{\psi}\psi \rangle_l/\partial m_l$ with the 
critical behavior $\chi_{m,l} \sim m_l^{1/\delta - 1}$. The $\chi_{m,l}$  is divergent in the chiral limit ($\delta>1$ for $O(N)$ scaling)
and the position of its peak is often used to define the critical (or crossover) temperature. An observable also sensitive to $f_s$,
but not widely used, is the mixed susceptibility $\chi_{t,l} = -  T\partial^2\ln Z/(V\partial m_l\partial t)$,
which has a divergent critical behavior: $\chi_{t,l} \sim m_l^{(\beta -1)/(\beta\delta)}$.

To define the critical temperature, sometimes so called "deconfinement type" observables are used, although their relation
to the $O(N)$ critical behavior is less clear or absent. One such observable is the quark number susceptibility
$\chi_{q=l,s}/T^2 = \partial^2\ln Z/(VT^3\partial(\mu_{q}/T)^2)$ which quantifies the
light or strange quark number fluctuations. The $\chi_{q=l,s}$ are expected to be zero at $T=0$, where the quark flavors
are confined in colorless states, and to increase as the temperature grows and the quark flavor is "freed".
The quark number susceptibilities are also important for the study of HIC since they are connected to event-by-event fluctuations of
various quantum numbers such as charge, strangeness and baryon number. For more details on how $\chi_{q=l,s}$
can be connected with various experimental observables see Ref.~\cite{mswagato}. The quark number susceptibilities
are somewhat sensitive to $f_s$ but they do not become divergent in the chiral limit. For example the critical behavior of the
temperature derivative of $\chi_{q}$ is:
\be
\frac{\partial \chi_q}{\partial T} \sim c_r +A_{\pm} \left| 
\frac{T-T_c^0}{T_c^0} \right|^{-\alpha},
\ee
where $\alpha<0$ for $O(N)$; hence, it is not divergent, and in fact, it is dominated by the regular part $c_r$.
Thus $O(N)$ scaling fits to the quark number susceptibility are not very reliable for extracting the critical
temperature.

Another often used deconfinement-type observable is the Polyakov loop which is related to the free energy of a static quark:
$\langle L \rangle\sim \exp(-F_Q/T)$.
In the limit of infinite quark masses it is governed by the singular part of the partition function of the pure gauge
theory, and it is an order parameter for deconfinement. At finite quark masses it is not, and its use as a mean to
define and extract $T_c$ has a less firm theoretical justification. Hence in this review I will concentrate
on results for the critical temperature obtained through only the chiral-type observables, which are
expected to be sensitive to the universal $O(N)$ scaling.
  
The extent to which universal scaling is applicable has been studied with both staggered \cite{stago4,abazavov} 
and Wilson \cite{wilo4} fermions. Generally it 
appears that the $O(N)$ scaling
gives an adequate description of the data.
At finite lattice spacing the scaling for rooted staggered fermions should be $O(2)$-type rather than $O(4)$, but in practice
the critical exponents for both types are similar enough that numerically they cannot be well distinguished \cite{stago4,abazavov}.
To illustrate this
statement, in Fig.~\ref{fig:diag} (right) scaling fits from Ref.~\cite{abazavov} to the renormalized order parameter
$M_b\equiv m_s \langle \bar{\psi}\psi \rangle_l/T^4=h^{1/\delta}f_G(z)$ are shown. 
Both $O(2)$ and $O(4)$ magnetic equation of state fits appear to perform similarly well. 
The resulting extrapolation to physical quark masses is also shown. 
If scaling is applicable, then by differentiating
these scaling fit curves with respect to the light quark mass, one should be able to predict the position of the peak of $\chi_{m,l}$.
In Fig.~\ref{fig:scaling} (left) a comparison between the $\chi_{m,l}$ data and the differentiated scaling curves is shown. 
Indeed, the predicted and measured positions of the peaks line-up satisfactorily. The scaling curves do not include regular parts
and this is the most likely reason for the discrepancy between the $\chi_{m,l}$ and the curves at lower $T$.  

\subsection{Critical temperature}
Since at physical quark masses there is no true phase transition to QGP at high $T$, the definition 
of a critical temperature $T_c$ is to some extent a matter of choice. However, a determination of  
$T_c$ based on observables sensitive to the singular part of the free energy $f_s$ is on a firmer theoretical 
footing and has a clearer interpretation. For this reason all of the results for $T_c$ quoted below are   
extracted from the peak in $\chi_{m,l}$.

Historically there appeared to be some discrepancies between $T_c$ values calculated for 2+1 flavors with different 
types of improved staggered fermions. In 2005 the MILC collaboration reported \cite{milctc} $T_c=169(12)(4)$ MeV
from a calculation with asqtad fermions. In 2006 the BNL-RBC-Bielefeld collaboration published \cite{bbrtc} the value $192(7)(4)$ MeV,
using the $p4$ action.
The Wuppertal-Budapest (WB) group in the period 2006-2010 consistently obtained \cite{wbtc} lower values with stout fermions, 
the latest of which is $147(2)(3)$ MeV. All of the above numbers are extrapolated to the continuum limit. 
The MILC result is also extrapolated to the chiral limit, while the other two values are at the physical quark mass limit.
The latest number from the HotQCD collaboration, obtained from calculations \cite{abazavov} with both the asqtad and the HISQ action is
$154(9)$ MeV (extrapolated to the continuum and physical quark mass limit). This result is compatible with the 
WB value. 
To illustrate this statement in Fig.~\ref{fig:scaling} (right)
the continuum
and physical quark mass extrapolations of the WB and HotQCD data for the subtracted chiral condensate 
\be
\Delta_{l,s} = (\pbp_{l,T} - \pbp_{s,T}\frac{m_l}{m_s})/(\pbp_{l,0} - \pbp_{s,0}\frac{m_l}{m_s})
\ee
are shown \cite{abazavov}. The stout and HISQ data have practically identical extrapolations,
suggesting the same $T_c$. As for the higher value of the BNL-RBC-Bielefeld result,
it is probably because the calculation was done at $N_t=4$ and $6$ only, which means relatively coarse lattices
were used and the discretization effects proved to be significant.

\begin{figure}[t]
\begin{tabular}{cc}
 \epsfxsize=62mm
  \epsfbox{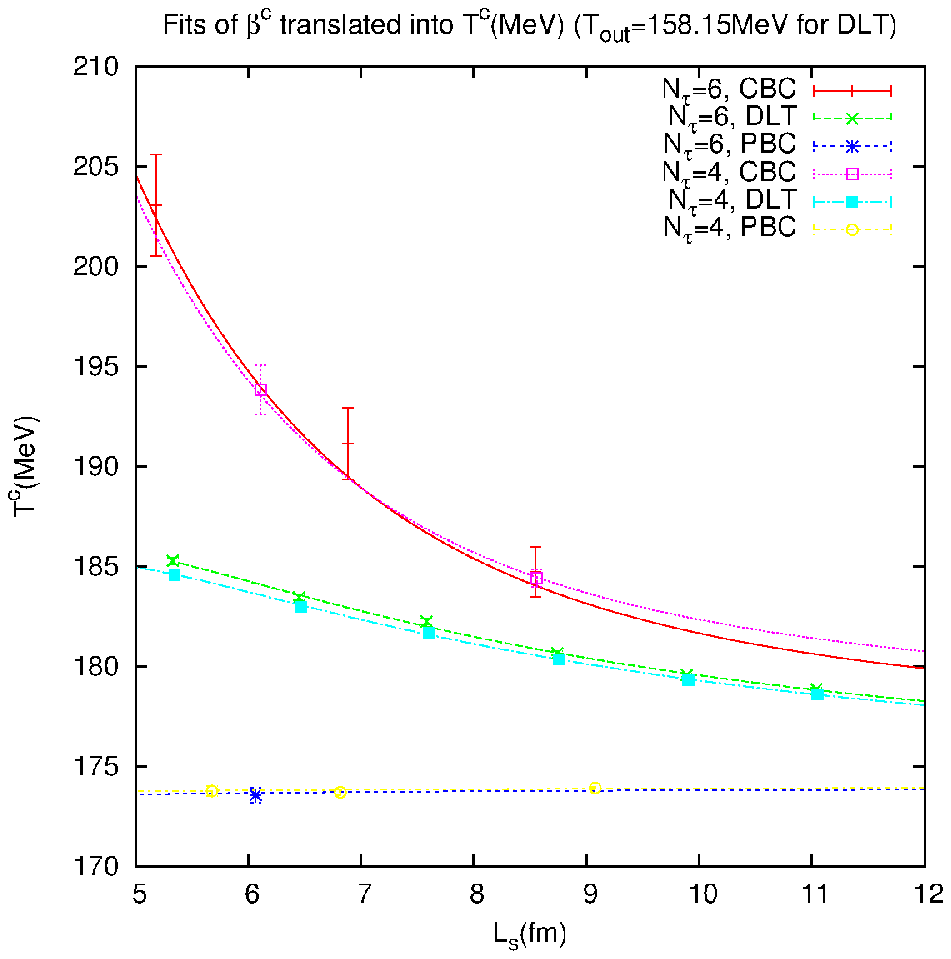}
&
  \epsfxsize=72mm
  \epsfbox{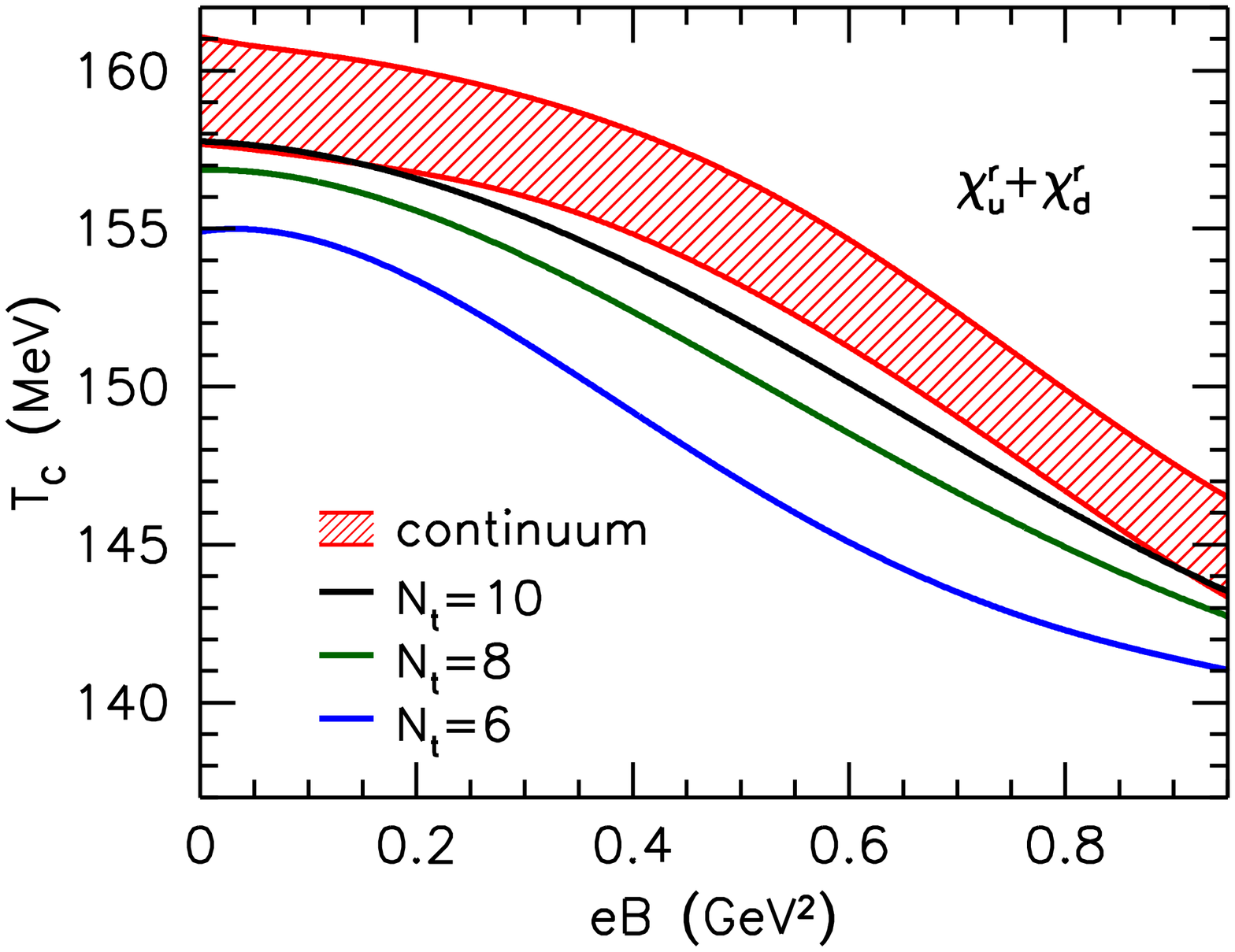}
\end{tabular}
\caption{(Left) The effects on $T_c$ from different lattice boundary conditions \cite{bberg}. (Right) The dependence of $T_c$
on the strength of the magnetic field $eB$ obtained from the peak of the average of the renormalized chiral
condensates $\chi^r_{u,d}$ \cite{gendrodi}.}
\label{fig:modif}
\end{figure}

All of the above results for $T_c$ have been obtained in the limit of "infinite" volume approximated through enforcing 
(anti)periodic boundary conditions (PBC) on the lattice. At HIC experiments the volume of the plasma created is only 5-10 fm$^3$ 
\cite{fv}.
The effect of such a small volume on the phase transition may be significant. In 2007 there was an attempt \cite{fve} to estimate it  
on the lattice by calculating in pure $SU(3)$ the effects of cold boundary conditions (CBC) on $T_c$ (meaning that
the ends of the lattice were left "free"). This setup is supposed to imitate a hot plasma immersed in a cold (zero temperature)
exterior. The results are shown in Fig.~\ref{fig:modif} (left) where the CBC are compared to the periodic
ones at different spatial lattice sizes $L_s$. It seems that the corrections to $T_c$ due to the finite volume effects
can reach up to 30 MeV. However, the CBC are not too easy to interpret, since the exterior temperature cannot be determined
straightforwardly. For $SU(3)$ in the scaling regime, the temperature can be set through the RG equation, but the cold exterior
has essentially zero coupling which is out of the range of its applicability. To address this difficulty this year an 
extended study \cite{bberg} was presented with a setup where both the interior and the exterior 
are kept in the scaling regime of $SU(3)$, through the
so called double-layered-torus (DLT) boundary conditions. The exterior has temperature of around 158 MeV, while the interior is
at $T_c$ (as determined from the Polyakov loops). Even with such a hot exterior by comparison with 
the HIC conditions, the $T_c$ with the DLT is still  
6-12 MeV higher than in the infinite volume case. It will be interesting
to know what the finite-volume correction would be with dynamical fermions added. 

Another factor that may influence the value of $T_c$ is the presence of strong external magnetic field. At HIC the 
magnetic field created in noncentral collisions is of $O(10^{14-15}\,\, \rm T)$ \cite{bhic}, a value much larger than a field of 
a typical magnetar $O(10^{10-11}\,\,\rm T)$ \cite{magnetar} , 
and not too far from the fields created at the beginning of the Universe ---  $O(10^{16}\,\,\rm T)$ \cite{bu}.
Strong external magnetic field may have a pronounced effect on the QCD phase transition, not only by modifying $T_c$
but also by changing its order, "separating" the chiral restoration from the deconfinement and inducing $CP$-violation 
through the Chiral Magnetic Effect (CME) \cite{cme}. Some effective models \cite{inctc} predict that $T_c$ will increase with stronger external
magnetic fields. Others find the opposite effect \cite{dectc}. It is important to settle these discrepancies with a first principle
lattice calculation. This year new results \cite{gendrodi} for the magnetic field effect on $T_c$ with stout fermions with physical masses\footnote{When it comes to staggered fermions, "physical" quark masses are defined as the ones which give
the lightest of the staggered pions and kaons their physical masses. It can be argued that since the rest of the
staggered multiplet members are heavier, the effective quark mass is actually larger than the physical even in this case.}
became available    
and are shown in Fig.~\ref{fig:modif} (right). The calculation was done at $N_t=6,8$ and $10$ and the continuum limit
was taken. The results show a decrease of $T_c$ when the magnetic field $eB$ increases, which is the opposite of what was found
in Ref.~\cite{latb}. However, the latter calculation was done at heavy quark masses (which the authors of Ref.~\cite{gendrodi}
claim as the most important factor for the difference), coarse lattices and with standard staggered
action, all of which may contribute to the discrepancy with this year's more advanced result. In both of these lattice calculations
the setup is such that the fermions interact with the external magnetic field but do not interact electromagnetically
among themselves. This "quenched EM" approximation introduces an unknown error in the final results. 
In the new study \cite{gendrodi} also no evidence was found for the transition becoming a 1st order in the range of the magnetic
fields available (up to around 1 GeV$^2$).  

\subsection{The equation of state}
\begin{figure}[t]
\begin{tabular}{cc}
 \epsfxsize=62mm
  \epsfbox{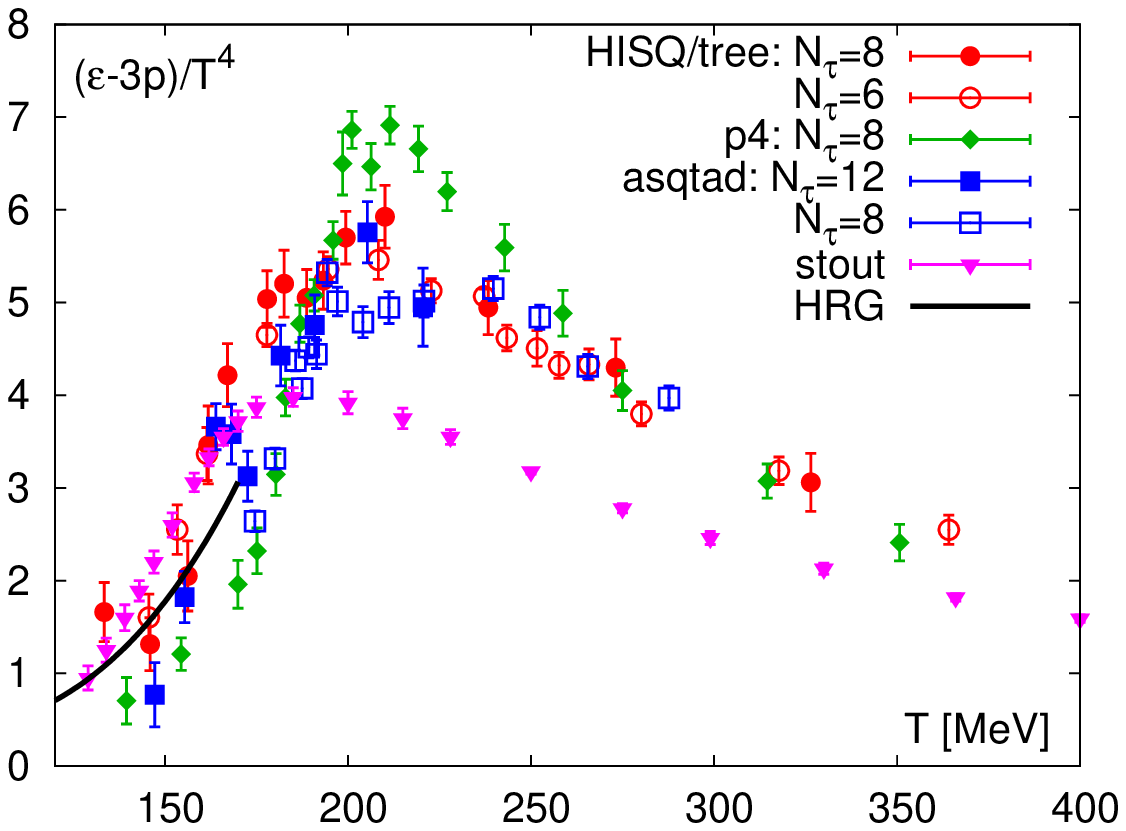}
&
  \epsfxsize=72mm
  \epsfbox{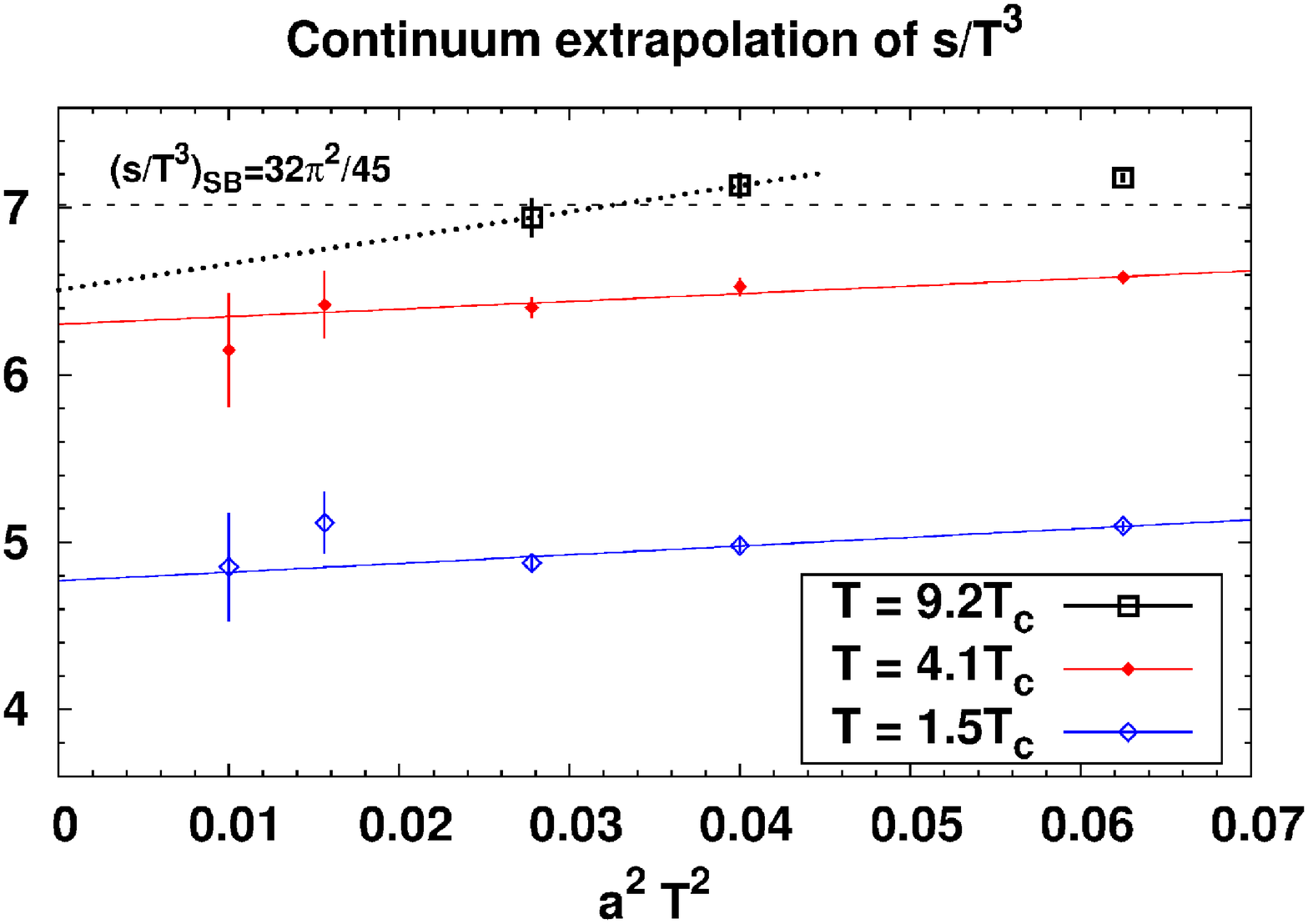}
\end{tabular}
\caption{(Left) The trace anomaly calculated with different staggered action and compared to
the HRG model at lower $T$ \cite{ws-2010}. (Right) The entropy density calculated at three 
values of $T$ with the new method from Ref.~\cite{G-M}. }
\label{fig:eos}
\end{figure}

The QCD equation of state (EOS) gives the temperature dependence of energy density, pressure and other bulk thermodynamic
quantities of a quark-gluon system.
It is an essential input to the hydrodynamic equations used to model the expansion 
of the plasma created at HIC. In recent years the demand for the EOS calculated from first principles
has grown \cite{bjacak}
as the HIC experiments accumulate high statistics on spectrum data. Lattice determinations of the EOS
are computationally expensive and  historically the most resources have been invested in the integral
method with variable scales. The integral method is based on the thermodynamic identities 
for energy density $\varepsilon$, pressure $p$ and the trace anomaly $\Theta^{\mu\mu}$:      
\be
 \varepsilon V = - \left.\frac{\partial \ln Z}{\partial(1/T)}\right|_V, \hspace{1cm}
  \frac{p}{T} = \left.\frac{\partial \ln Z}{\partial V}\right|_T \approx \frac{\ln Z}{V}, \hspace{1cm}
   \Theta^{\mu\mu} = \varepsilon - 3p = -\frac{T}{V} \frac{d \ln Z}{d \ln a}.
\label{eq:eos}
\ee
From the above it follows that $p$ and $\varepsilon$ can be obtained by integrating $\Theta^{\mu\mu}$ over the lattice spacing 
(hence
the method's name). In the most widely used version of this method \cite{milceos}, the lattice spacing is varied along an LCP
in order to change $T$ while $N_t$ is kept fixed. This means that  at lower temperatures the lattices
are coarser and the discretization effects larger. In the fixed-scale version of the integral method \cite{fixed},
where $a$ is kept constant and $N_t$ varies (and the integration is over $N_t$), the situation is reversed ---
the larger discretization effects will be at high temperatures. There is another fixed-scale method which
does not require integration, called "the operator/differentiation method" \cite{oper}. The pressure and energy density
in this case are calculated directly from the first and second expressions in  Eq.~(\ref{eq:eos}) (i.e.,   
not through
the determination of $\Theta^{\mu\mu}$ first).
However, the method demands the determination of anisotropy coefficients in addition to the beta- and
mass-renormalization functions required for the integral methods. All of the methods described require a zero temperature
subtraction to eliminate the UV divergences in the EOS.
\begin{figure}[t]
\begin{tabular}{cc}
 \epsfxsize=62mm
  \epsfbox{p_charm.eps}
&
  \epsfxsize=72mm
  \epsfbox{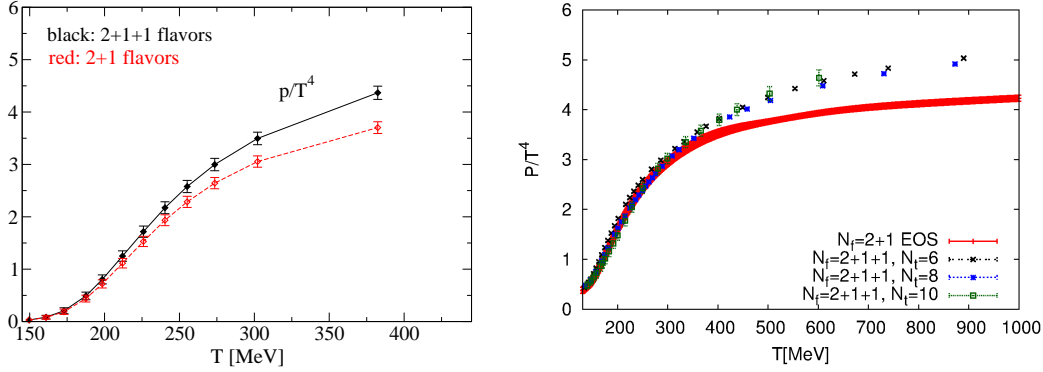}
\end{tabular}
\caption{(Left) The pressure obtained with the heavy-quark quenched approximation in the 2+1+1 flavor case and 
compared with the 2+1 (dynamical) flavor result \cite{hqq-mine}. (Right) The pressure calculated with 2+1+1 dynamical flavors
compared with the 2+1 flavors \cite{skrieg} . }
\label{fig:c_eos}
\end{figure}

In Fig.~\ref{fig:eos} (left) the values for the trace anomaly are compared for several staggered actions and at different
$N_t$'s. The method of calculations is the variable scale one. The data for the asqtad, $p4$ and the HISQ action
is from the HotQCD collaboration \cite{ws-2010}. It is calculated on an LCP, where $m_s$ is physical and $m_l/m_s=0.05$.
The stout data is from the WB group \cite{wb-eos} and is obtained with all quark masses set to the physical ones and at
$N_t=8$ (they also have data for $N_t=10$ and 12 \cite{wb-eos,skrieg} which agree with the $N_t =8$ case).
At lower temperatures all the results are compared with the hadron resonance gas (HRG) values, which
is expected to be a reasonable description of the system in this regime \cite{hrg}. The data at larger $N_t$ (finer lattices)
start to approach the HRG result at low $T$ and the small difference which remains between 
the WB and HotQCD data there may be attributed to the difference in the quark masses. The more striking difference
in the region above the crossover is still without explanation. However it is probably not a result of 
quark-mass differences (or deviations from the LCP), since these should play a minor role at high $T$.
It is possible that additional HISQ runs at $N_t=12$ may help to explain this discrepancy.

For HIC at RHIC the EOS at 2+1 flavors is considered most relevant, since the time scales are probably too 
short for the thermalization of the 
charm quark \cite{laine}. At LHC the quark-gluon plasma is hotter and lasts longer and thus it is possible
that the charm quark thermalizes in this case. This implies that its contribution to the EOS of the
plasma created at LHC
may not be negligible. Previous heavy-quark quenched calculations \cite{hqq,hqq-mine} (i.e., where the sea charm loops are neglected)
attempted to calculate the effects
of the charm quark (see Fig.~\ref{fig:c_eos} (left) for pressure results). But the question remained
whether the quenched approximation used for the charm quark introduces a substantial systematic error. A new study \cite{skrieg}
with stout fermions at $N_t=6,8$ and 10, where all the $u,d,s$ and $c$ flavors are dynamical,
points toward
an answer to this problem. In Fig.~\ref{fig:c_eos} (right) the results for the pressure for 2+1+1 dynamical
flavors is presented. By comparison with the heavy-quark quenched result from Ref.~\cite{hqq-mine}
in the left panel of the same figure,
the contribution of the charm quark is smaller throughout and becomes significant only at relatively
larger temperatures
($T>350$ MeV). The authors also show that 
dynamical $c$-quark effect in the pressure  is very close to 
its perturbative estimate \cite{laine-sch}. They attribute the difference between the quenched and dynamical cases
to a shift in the LCP which results from the $c$-quark being part of the sea. 
There still remains the question of the heavy-quark discretization effects and how well the charm quarks are
represented on the lattices used in Ref.~\cite{skrieg} with stout fermions.
\begin{figure}[t]
\begin{tabular}{cc}
 \epsfxsize=60mm
  \epsfbox{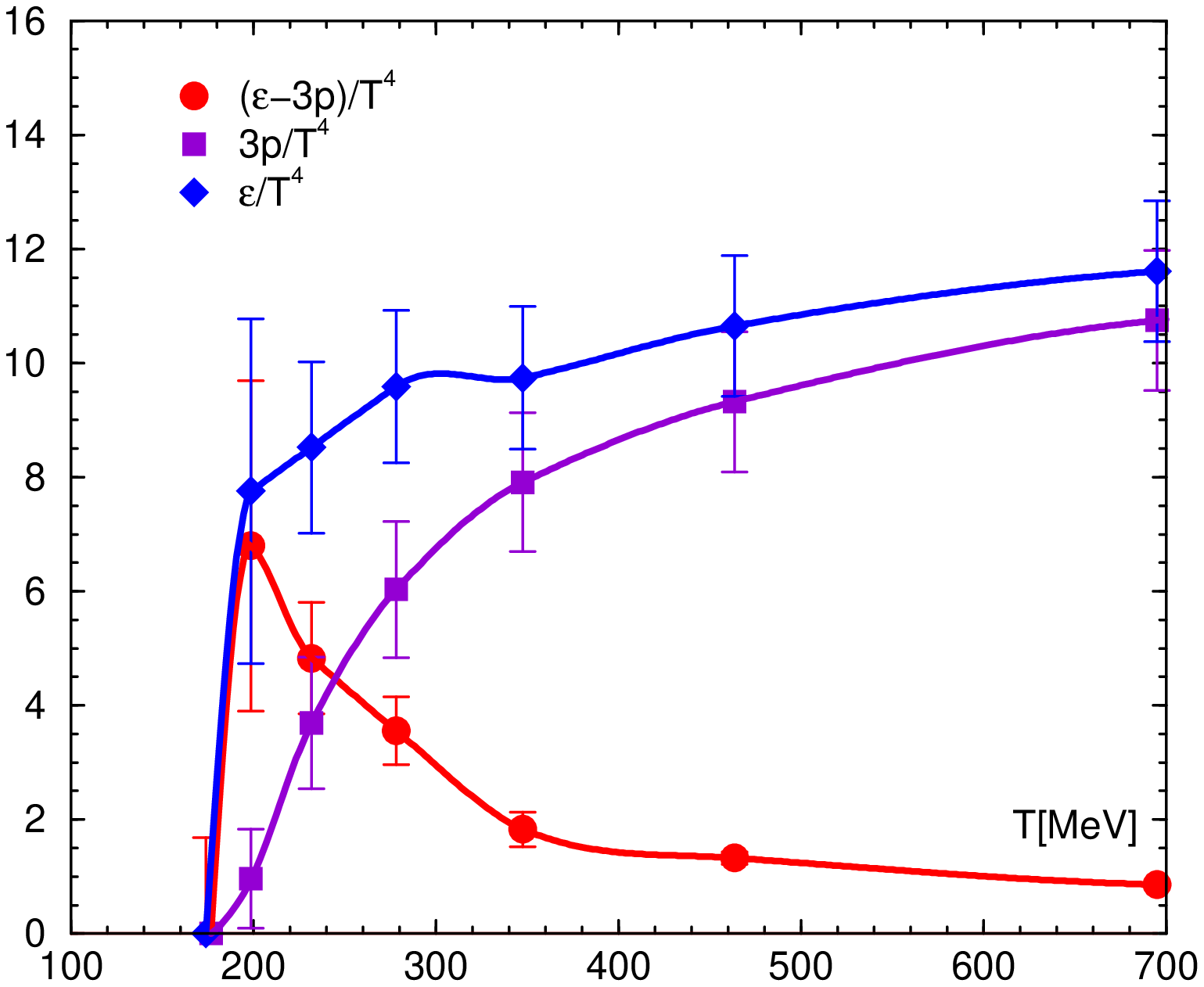}
&
  \epsfxsize=72mm
  \epsfbox{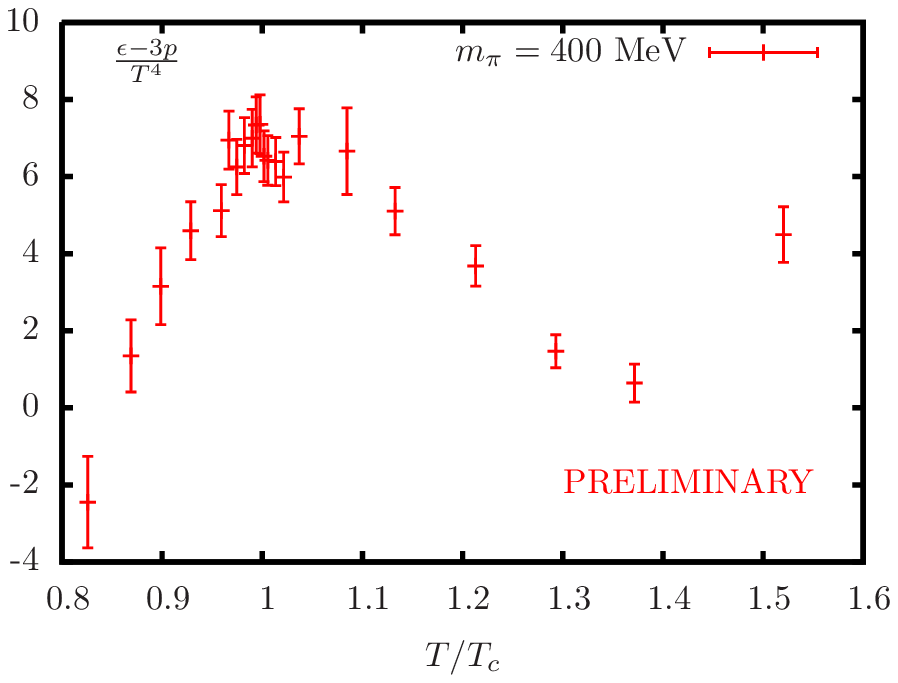}
\end{tabular}
\caption{(Left)WHOT-QCD EOS result \cite{fixed} for 2+1 flavors of nonperturbatively improved Wilson fermions.
The lattice scale is fixed to $a\approx 0.07$ fm and $N_t$ is varied between 4 and 16 and $m_\pi/m_\rho\approx0.63$. (Right)
New result \cite{fburger} for the trace anomaly calculated with 2 flavors of mtmWilson fermions at $N_t=12$. The lattice scale is varied $0.04-0.1$ fm. The outliers are probably due to finite volume effects and zero-temperature interpolations. }
\label{fig:wilson_eos}
\end{figure}

The EOS is also studied with Wilson-type fermions; Fig.~\ref{fig:wilson_eos} shows two of the most recent results. 
The WHOT-QCD
collaboration used 
the integral method with fixed scales and the data from Ref.~\cite{fixed} is shown in the left panel of this figure.
In the right panel of Fig.~\ref{fig:wilson_eos} a new result 
\cite{fburger} with mtmWilson fermions is presented.
In this case the method with variable scales is employed. More details of these two calculations
are given in the caption of Fig.~\ref{fig:wilson_eos}. Both of these studies are at heavy quark masses and 
have large statistical errors, and for the time-being, they are not yet as sophisticated as the staggered calculations. However,  
Wilson studies are essential for verifying the current
staggered results with an alternative action which does not suffer from taste symmetry violations. This year
such a staggered-Wilson comparison was reported for some thermodynamic quantities \cite{nogradi} (namely: $\pbp$, $\chi_s$ and the Polyakov loop), which show a   
very good consistency between the two different types of fermion action.

A qualitatively new method for calculating bulk thermodynamic quantities was recently introduced in Ref.~\cite{G-M}.
It is based on a lattice determination of the generation function of the cumulants of the momentum distribution
$K(\beta,\mu,z)$. The cumulants of $K(\beta,\mu,z)$, where $z$ is a 3 component vector, defined as 
\be
K_{2n_1,2n_2,2n_3}=(-1)^{n_1+n_2+n_3+1}
\frac{\partial^{2n_1}}{\partial z_1^{2n_1}}\frac{\partial^{2n_2}}{\partial z_2^{2n_2}}
\frac{\partial^{2n_3}}{\partial z_3^{2n_3}}\frac{K(\beta,\mu,z)}{N_s^3}|_{z=0}\quad ,
\ee
are related to the entropy density $s$ and
the heat capacity $c_\nu$ in the following manner:
\be
K_{2,0,0} = T^2s,\;\;\;\;\;\;\;\; \frac{K_{2,2,0}}{T^4}- \frac{3K_{2,0,0}}{T^2}=c_\nu.
\ee
The above means that the knowledge of $K(\beta,\mu,z)$ would give access to the
determination of the EOS, since by integrating $s$ over the temperature one can
obtain the pressure using the identity $s =T dp/dT$. 
To calculate $K(\beta,\mu,z)$ the following result from thermal quantum field theory is employed:
\be
e^{-K(\beta,\mu,z)}=\frac{Z(\beta,\mu,z)}{Z(\beta,\mu)},
\label{eq:ratio}
\ee
where $Z(\beta,\mu,z)=\Tr\{e^{-\beta(\hat{H}-\mu\hat{N})-i\hat{p}z}\}$ is a partition function where
all states with momentum $p$ acquire a phase $e^{ip.z}$. On the lattice this partition function can
be represented as a Euclidean path integral with shifted b.c. for the fields in the temporal direction:
$\phi(t,x)=\pm\phi(0,x+z)$. According to the method in Ref.~\cite{G-M} applied at 
$\mu=0$, to evaluate the partition function
ratio in Eq.~(\ref{eq:ratio}) a system of $N$ interpolating ensembles with \maths{Z(\beta, r_i)}
is created, each with the following
action:
\be
\overline S(U,r_i)= r_i S(U) + (1-r_i) S(U^z),
\ee
where $S(U)$ is an action with PBC, $S(U^z)$ is one with shifted b.c. and $r_i=i/N,\,i=0,\dots,N$.
To have the discretization effects under control the number $N$ should be $O(N_s^3)$.
Then \maths{K} is approximated as:
\be
K(\beta,z,a)=-\ln\frac{Z(\beta,z)}{Z(\beta)} \approx -\sum_{i=0}^{N-1}\ln
\frac{Z(\beta,r_i)}{ Z(\beta,r_{i+1})}=-\sum_{i=0}^{N-1}
\ln\langle e^{{\overline S}(U,r_{i+1})-{\overline S}(U,r_i)}
\rangle_{r_i+1}.
\ee
The derivative of $K(\beta,\mu,z)$ to determine the entropy is taken as the following limit:
\be
\frac{s(T)}{T^3} = \frac{K_{2,0,0}(\beta)}{T^5} = 
\lim_{a\rightarrow 0} 
\frac{2 K(\beta, z, a)}{{|z|}^2 T^5 N_s^3}.
\ee
The deviation from the continuum at finite lattice spacing with the above procedure is $O(a^2)$.
The authors of Ref.~\cite{G-M} have applied the method to the pure $SU(3)$ case using a plaquette action. 
In Fig.~\ref{fig:eos} (right) the results for $s$ on the lattice spacing are shown for three 
temperatures. The approach to the continuum at the lower two $T$'s seems rather mild. At the
highest available $T$ the discretization effects seem more pronounced, probably due to 
the lattice spacings 
being somewhat coarse for $T=9.2T_c$ and the UV cutoff being more pronounced in this case.
The computational cost for this new method could be as high as $O(a^{-11})$
for the pure gauge theory depending on the algorithm for estimating the ratio in Eq.~(\ref{eq:ratio}).
The method does not require zero temperature subtractions
to eliminate the UV divergences in thermodynamic quantities, but the overall cost is likely not 
going to be lower than the one
for the more traditional methods. 
  
\section{QCD at nonzero baryon density}
\begin{figure}[t]
\begin{tabular}{cc}
 \epsfxsize=72mm
  \epsfbox{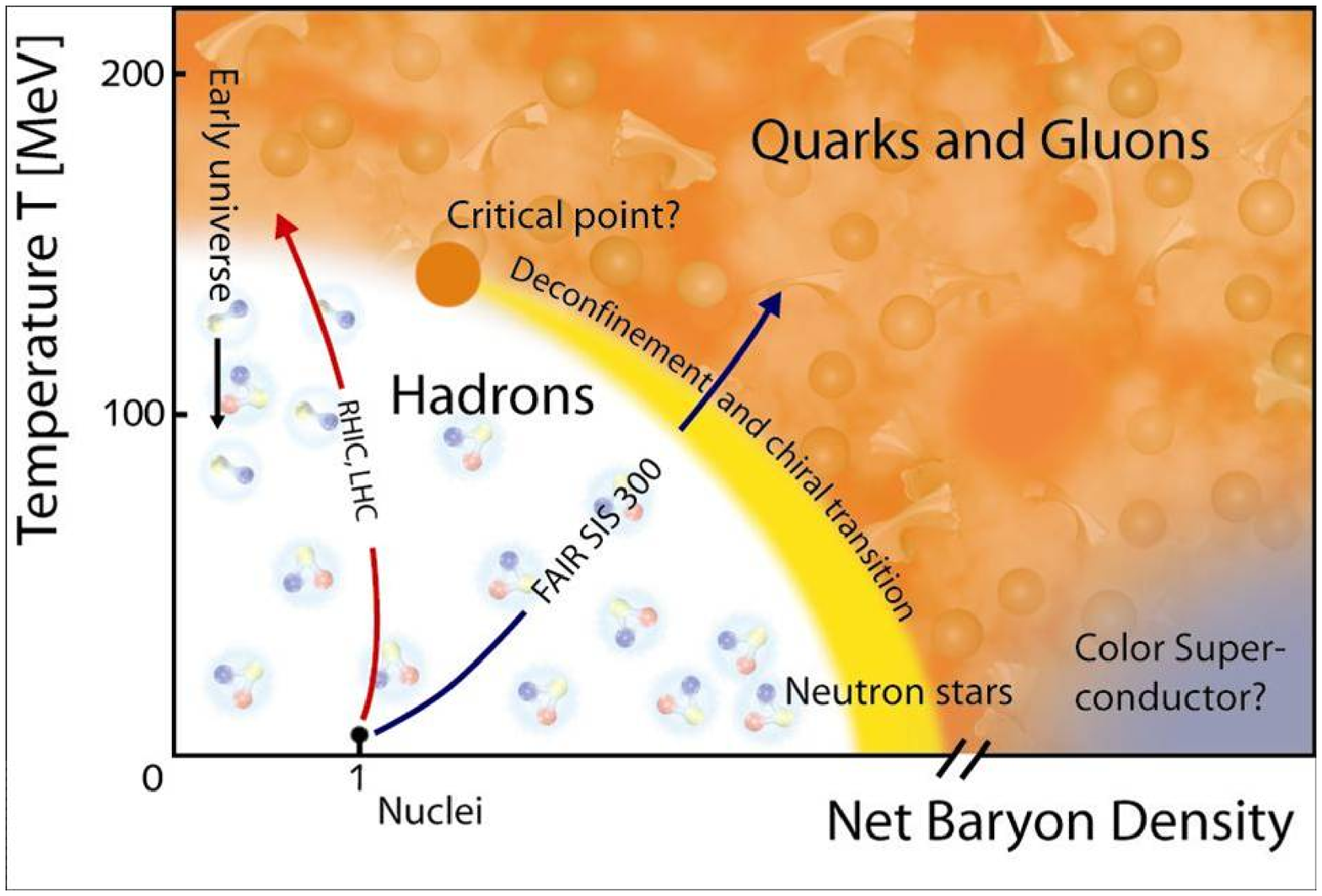}
&
  \epsfxsize=72mm
  \epsfbox{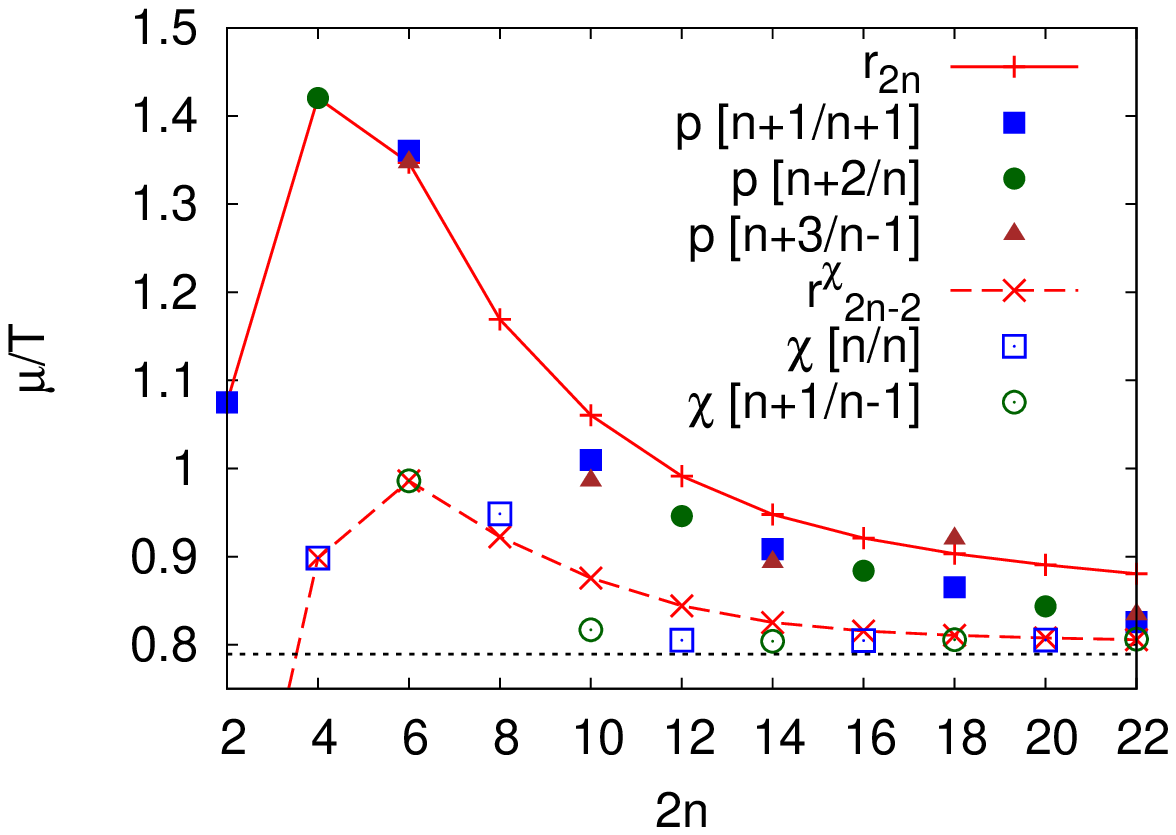}
\end{tabular}
\caption{(Left) The QCD phase diagram at nonzero baryon density. (Right) Estimate of the 
phase boundary at $T=190$ MeV
(which is slightly higher than $T_E$) as a function of
the order of the expansion
from $r_{2n}$, $r_{2n}^\chi$ and Pad\'e approximations of order $p/q$ \cite{pqm1}.
The dotted line is the full model solution.} 
\label{fig:muT}
\end{figure}
Studying QCD at nonzero baryon density (or nonzero chemical potential $\mu$) is important 
for a variety of physical phenomena
such as supernova explosions, neutron star formation, the existence of quark stars and degenerate matter.
The properties of hot hadronic matter at nonzero baryon density are also explored experimentally
at lower-energy HIC. 
Thus a quantitative determination of the QCD phase diagram in the $T-$baryon density plane is of both great theoretical and experimental
interest. The current conjecture for the diagram is shown in Fig.~\ref{fig:muT} (left). At high $T$ and low
density it is accepted that the transition is a crossover. It is expected that as the baryon 
density increases the
crossover region moves to lower $T$ until a 2nd order critical end point (CEP) is reached. At even
higher densities the transition is supposed to become of 1st order. On the diagram the first
order region is denoted by a band since for a 1st order transition the two phases coexist 
at a (narrow) range of baryon densities at the critical temperature. This first order 
band shrinks to a line when the
diagram is presented in the $T-\mu$ plane instead (i.e., the diagram in Fig.~\ref{fig:muT} (left)
is from a canonical ensemble point of view, instead of the grand canonical one in the $T-\mu$ plane).
It is also important to establish quantitatively the relative position of the experimental
freezeout curve \cite{freeze} with respect to the critical one, since if the two are far apart the measured signals
at HIC experiments for the 1st order phase transition may become "washed out" and more difficult to interpret. 

\subsection{Nonzero chemical potential on the lattice}
Introducing nonzero chemical potential on the lattice presents a significant challenge --- 
the straightforward Monte Carlo (MC) simulations, where the determinant of the fermion matrix
$\det M$ is interpreted as a probability density, become unfeasible. The reason for this
is that $\det M(\mu\neq0)$ becomes complex:
\be
(\det M(\mu))^\star=\det M(-\mu)\neq \det M(\mu).
\ee
This constrains the MC simulations to $\mu$ being zero, or pure imaginary, or to
the case of pair(s) of degenerate flavors with opposite real $\mu$ (nonzero isospin chemical potential)
since in those cases $\det M$ remains real.
There are a host of methods which attempt to avoid the problem of the complex determinant 
(also referred as "the sign problem") at $\mu\neq0$. Examples are: the Taylor expansion method \cite{taylor},
reweighting techniques \cite{rew}, analytic 
continuation from imaginary $\mu$ \cite{imu}, density of states methods \cite{dos}
and canonical ensemble approaches \cite{canon}. 
All of these methods find a way to circumvent directly simulating at real nonzero $\mu$.
In the next subsection I discuss in more detail the Taylor expansion method only. For 
recent reviews of the rest of the methods see Ref.~\cite{rev,rev1}.

There are two methods "under construction" which bear some promise to be suitable for 
direct QCD simulations at any $\mu\neq0$ in the future. The first one is the method
of stochastic quantization \cite{stoch}, where the evolution of the gauge fields is governed by 
complexified Langevin equations, which can be integrated even at nonzero $\mu$.
In practice the method is prone to instabilities and still at the stage where
the conditions for the convergence to the correct ensemble are explored. This year
results \cite{fjames} from testing the method in an $SU(3)$ spin model, which approximates the 
strong-coupling large-quark-mass regime of QCD, show that the method gives the
correct solution for this particular case. The application of stochastic quantization 
also was studied
in the chiral random matrix theory at finite density \cite{tsano}, where analytical and numerical
solutions can be compared.
A second method for direct simulation and nonzero density is the world-line approach \cite{wline},
the basic idea of which is to integrate out explicitly the gauge fields instead of the 
fermion ones as
is usually done. This way the sign problem could be alleviated \cite{wline1}. Unfortunately the gauge field
integration cannot be performed explicitly due to the four field interactions, except in the
strong coupling limit of QCD. In this limit after gauge integration the remaining degrees
of freedom are mesons and baryons i.e., color singlets. The "worm" algorithm \cite{worm} is one best
suited for this setup and this year a new "continuous time" version \cite{wunger} of it was introduced which does not 
suffer from the sign problem and was applied to the study of the phase diagram in 
the strong coupling regime of QCD.  Both of these methods (stochastic quantization and world-line
approach) were reviewed  previously in more detail \cite{rev}. 

\subsection{The Taylor expansion method}
The main idea of the Taylor expansion method is that the basic thermodynamic
quantities are represented as series whose coefficients can be calculated
on an ensemble with $\mu=0$ where a traditional MC simulation is possible, which is the method's 
most important advantage. In other words,
the expansion is around $\mu=0$ and the expansion parameter is $\mu/T$. 
For the pressure the expansion explicitly is:
\be
{p(T,\mu)\over T^4}={\ln Z \over VT^3}=
\sum_{n=0}^\infty c_{n}(T) \left({\mu\over T}\right)^n ,\hspace{1cm}
c_n(T)= \left.\frac{1}{n!VT^3}\frac{\partial^n\ln Z}{\partial(\mu/T)^n}\right|_{\mu=0}\, ,
\ee
where $\mu$ is the light quark chemical potential and we assume the heavy quark
one is zero for simplicity.
\begin{figure}[t]
\begin{tabular}{cc}
 \epsfxsize=72mm
  \epsfbox{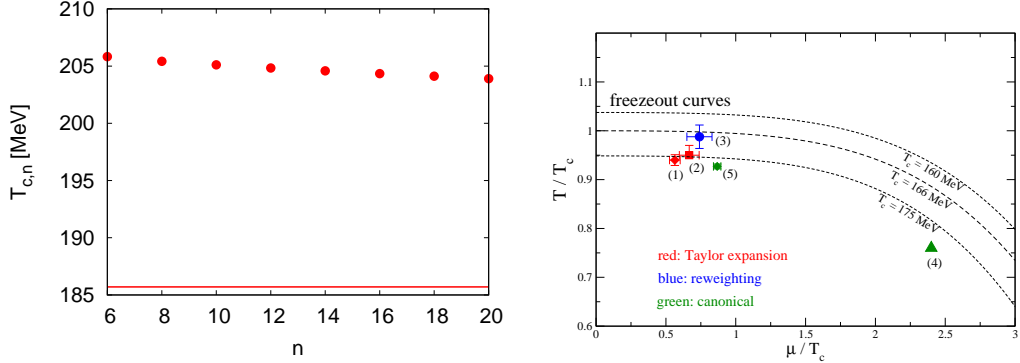}
&
  \epsfxsize=62mm
  \epsfbox{CEP.eps}
\end{tabular}
\caption{(Left) The CEP temperature estimates  from the first max of \maths{c_n} \cite{pqm1}.
Solid line is the full model $T_E$.  (Right) The CEP location determined with different
methods and compared to possible positions of the freezeout curve. Taylor expansion method
(1): standard staggered $N_f=2$, 
$N_t=6$, volumes up to $V=24^3$, $m_\pi/m_\rho\approx0.3$ \cite{gagu}; (2): $p4\,\,$ $N_f=2+1$, $N_t=4$, volumes
up to $V=24^3$, $m_\pi/m_\rho\approx0.3$ \cite{cep2}.  Reweighting (3): staggered $N_f=2+1$, $N_t=4$, 
volumes up to $V=12^3$, $m_\pi/m_\rho=\,\rm physical$ \cite{cep3}. Canonical approach (4): $p4$ $N_f=2$, $N_t=4$,
volume $V=6^3$, $m_\pi/m_\rho\approx 0.7$ \cite{cep4}. Direct canonical simulation (5): clover $N_f=3$, 
$N_t=4,\, V=6^3$, 
$m_\pi=700-800\,\rm MeV$ \cite{cep5}.}
\label{fig:cep}
\end{figure}

The expansion coefficients $c_n(T)$ are nonzero only if $n$ is even due to the CP
symmetry of the partition function. The expansion is roughly expected to converge
when $\mu$ is smaller than the first Matsubara frequency i.e., when 
$\mu\lsim \pi T$. However, in the chiral limit the radius of convergence of the Taylor expansion will shrink
to zero at the CEP. Even at nonzero quark masses and at small $\mu$, in the crossover
region the statistical fluctuations in the Taylor coefficients grow large and make them
more difficult to determine. The method has also other drawbacks which should be taken 
cautiously into account.
For example, the number of terms in a given $c_n$ grows approximately as $6^n$ and
there are large cancellations between them. The $c_n$'s are estimated stochastically and 
the number of noise vectors required to keep the noise under control grows exponentially  
with $n$ and the volume \cite{rev}. As a whole the computational effort grows factorially with
increasing order of the expansion, the highest reached for QCD calculation being $n=8$ \cite{gagu}.
Higher orders are also sensitive to finite volume effects since essentially $c_n$
is an $n$-point correlation function.

The position of the CEP in principle can be estimated using the Taylor expansion method. The quark number susceptibility, whose expansion is
\be
\frac{\chi_q(T,\mu)}{T^2} = \frac{\partial^2 p(T,\mu)}{\partial (\mu/T)^2}=\sum_{n=2}^{\infty}
  \underbrace{n (n-1) c_n(T)}_{ =c^\chi_{n-2}(T)}
  \left(\frac{\mu}{T}\right)^{n-2},
\ee
is expected to diverge at the CEP (unlike the pressure which stays continuous). 
The  
radius of convergence of the series for the pressure and the $\chi_q(T,\mu)$ is the same
 and can be related to the coordinates of the CEP ($\mu_E,\,T_E$) if the following
condition is fulfilled. If in the complex plane the closest singularity of the expansion happens 
at real $\mu$ then the radius of convergence $r$ is related to the CEP coordinates as
\be
r = \mu_E/T_E = \lim_{n \rightarrow \infty}
  \left|\frac{c_{2n}}{c_{2n+2}} \right|^{1/2} = \lim_{n \rightarrow \infty}\left|\frac{c^\chi_{2n}}{c^\chi_{2n+2}} \right|^{1/2}  .
\label{eq:r}
\ee
Mathematically, for the singularity to lie on the real axis there should exist an order $n_0$ such that
for each $n>n_0$, all $c_n>0$. 
 
An estimate for $T_E$ can be obtained independently of $r$ from the temperature behavior of the coefficients
$c_n$. Close to the CEP a derivative of $c_n$ with respect to the temperature is equivalent 
to two derivatives with respect to $\mu$, essentially 
meaning that $\partial c_n/\partial T\sim c_{n+2}$ \cite{pqm1}.
This property can help in the determination of the temperature around which all coefficients with
order higher than $n$ are positive, by
finding the temperature of the first maximum of $c_n$. This gives consecutive estimates for $T_E$
as $n$ is increased which are expected to converge from above to the CEP temperature. From estimates
of $r$ and $T_E$ the critical chemical potential $\mu_E$ can also be found through Eq.~(\ref{eq:r}).
The most important question about the Taylor expansion method is:
how many orders are needed for a reliable determination of the CEP coordinates (or other
thermodynamic quantities)? The answer to 
this question through direct QCD simulations is difficult due to the large computational 
resources required to
reach higher orders. Instead some insights into this question may come from model calculations
for which these type of computations are much cheaper.

The PQM model, whose degrees of freedom are Polyakov loops, quarks and mesons, has been shown to
be a good approximation of QCD for 2+1 flavors and at $\mu=0$ when solved in a mean field approximation \cite{pqm}.
This year the same model has been explored (still in the mean field approximation) in the case
of $\mu\neq0$ \cite{pqm1}. The authors compare the mean-field full solution results with ones obtained
from a Taylor expansion to $O(22)$. In Fig.~\ref{fig:muT} (right) shown is the 
result for the distance to the
phase boundary from successive estimates of the convergence radius, which is determined both  
from the coefficients of the pressure expansion and from the quark number susceptibility one
(see Eq.~(\ref{eq:r})).
Also shown are data points for the first pole in Pad\'e approximations to the pressure and    
$\chi_q(T,\mu)$, which is an alternative method for finding the distance to the phase boundary.
From the figure the following conclusion can be made: the estimates $r_{2n-2}^{\chi}$ 
converge faster to the full solution than the $r_{2n}$ ones, and the fastest convergence
is achieved through Pad\'e  approximation of $\chi_q(T,\mu)$. It appears that at least $O(10)$
in the Taylor expansion of $\chi_q(T,\mu)$ is needed in order for the result to be within
10\% of the full solution one. In Fig.~\ref{fig:cep} (left) shown are the data for the estimate for 
the temperature of the critical end point $T_{c,n}$,
as a function of the order of the coefficient $c_n$, from whose maximum it has been determined.
From the figure it follows that the approach to the full solution $T_E$ is very slow with this method
and even for $O(20)$ the estimate is still about 10\% off, which can 
lead to a significant systematic error in the determination of $\mu_E$. This study may point to a 
potential problem with slow convergence of the Taylor expansion for the true QCD case,
to the extent the model is a relevant description for it. It is also possible that  
the slow convergence is a feature of the model rather than the real QCD case, meaning that a particular model's
behavior may only be appropriate just as a guide.
Still, another recently studied QCD model \cite{ising} (3d Ising model) estimates that
at least $O(16)$ of the Taylor expansion is needed, in support of the claim that the inclusion of 
high orders may be necessary.

\subsection{The critical end point and the curvature of the critical line}
\begin{figure}[t]
\begin{tabular}{cc}
 \epsfxsize=72mm
  \epsfbox{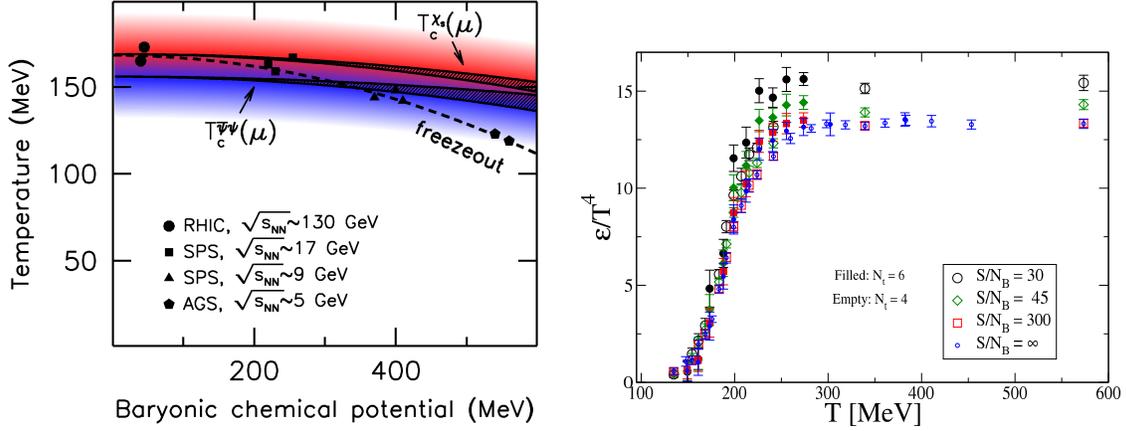}
&
  \epsfxsize=72mm
  \epsfbox{EO6Isentr.eps}
\end{tabular}
\caption{(Left) The critical line as determined in Ref.~\cite{k} and its position
with respect to the freezeout curve. (Right) The isentropic energy density for different
constant values of the ratio of the entropy $S$ and baryon number $N_B$ \cite{hqq-mine}.} 
\label{fig:kappa}
\end{figure}
In this subsection I review some recent quantitative results for the features of the QCD phase diagram 
at nonzero chemical potential. 
For example, the location of the CEP has been the object of a number of studies using variety of methods
\cite{gagu,cep2,cep3,cep4,cep5}.
In Fig.~\ref{fig:cep} (right) a number of recent values for it are shown. There are two results
obtained from the Taylor expansion method as described in the previous subsection, denoted by
(1) and (2). One of the results (3) is from reweighting techniques and the rest are 
from canonical approaches (points 4 and 5). The results appear to cluster between $\mu/T_c=0.5-1.0$
and $T/T_c=0.9-1.0$, except one of them (4). Also shown in this figure are a host of possible
freezeout curves which differ only on the assumed value of $T_c$ to which they are normalized.
It appears that the cluster of CEP results lies close to the freezeout curves. (However,  
the results for  
the CEP location
from the analytic continuation method \cite{analyticCEP} are in a 
qualitative disagreement with all of the above. They are 
interpreted to mean that the CEP may not
even exist at physical quark masses.)  

Another important feature of the diagram is the curvature of the critical line,
and more interestingly, its comparison to the one of the experimentally determined
freezeout curve. The critical line curvature $\kappa$ at $\mu=0$ can be determined using the
Taylor expansion method where to lowest order
\be
T_c(\mu_B^2)=T_c \left( 1 - \kappa \cdot \mu_B^2/T_c^2 \right),\hspace{1cm}\kappa= - T_c \left. \frac{d T_c(\mu_B^2)}{d (\mu_B^2)} \right|_{\mu_B=0},
\ee  
with $\mu_B=3\mu$ being the baryon chemical potential. In a recent work \cite{k} $\kappa$ has been determined
through the following expression:
\be
\kappa=  -T_c\left[- \left.\left( \frac{\partial \phi}{\partial \mu_B^2} \right)\right|_{T_c,\mu_B=0}\Big /\left.\left(\frac{\partial \phi}{\partial T} \right)\right|_{T_c,\mu_B=0}\right],
\ee
where $\phi$ is an auxiliary observable, for example $\pbp$ or $\chi_q$, the derivatives
of which are calculated at the critical
temperature and on an ensemble at $\mu_B=0$. This study with 2+1 flavors of stout
fermions was done at $N_t=6,8$ 
and 10 with physical
quark masses and the continuum
limit of the data was taken. The results for the curvature are $\kappa^{(\chi_s/T^2)} = 0.0089(14)$
and $\kappa^{(\bar\psi\psi_r)} = 0.0066(20)$ from data for the strange quark number susceptibility
and the chiral condensate. In Fig.~\ref{fig:kappa} (left) the resulting
critical lines (following from both observables) are shown at small $\mu_B$ assuming that the curvature
does not depend on the temperature. The two values for $\kappa$ are compatible within their errors and are also
broadly compatible with results from other studies. For example, a p4 study with 2+1 flavors
gave $\kappa = 0.0066(5)$ using $O(4)$ scaling analysis \cite{k1}. Results with two flavors also appear close
to these values: $\kappa = 0.0059(1)$ from analytical continuation with staggered fermions \cite{k2}, and 
$\kappa\approx0.0078$
from $O(4)$ scaling analysis with Wilson fermions \cite{wilo4}c. This closeness between the 2 and 2+1 flavor results is based on the expected
similarity of the critical behavior in both cases. The curvature of the freezout curve at $\mu_B=0$ is
experimentally estimated to be $\kappa\approx0.02$ \cite{freeze}, which is a few times larger than the estimated
values for the critical line one, meaning that, provided the curvature of the critical line does not change with $T$,
the two curves will increasingly diverge as $\mu_B$ 
grows. 
 
\subsection{The EOS at nonzero chemical potential}   

The EOS at nonzero $\mu$ is important for HIC experiments when the center of mass energy is
low, in which case the baryon density grows. More accurately, the relevant EOS is the isentropic one,
where the ratio of the entropy and the baryon number ($S/N_B$) is constant. 
In Fig.~\ref{fig:kappa} (right) results for the energy density at different $S/N_B$ from
one of the most recent studies \cite{hqq-mine} are shown.
The values for $S/N_B$ are chosen such that they correspond to the ones achieved at different
experiments: 30, 45 and 300 are for AGS, SPS and RHIC, respectively. The calculation was done
with the Taylor expansion method to $O(6)$, 2+1 flavors of asqtad fermions at $N_t=4$ and $6$ and $m_l/m_s=0.1$.
As the temperature
is changed the light and heavy quark chemical potentials were tuned such that the
strange quark number density is zero (within statistical errors). The results show that
for large $S/N_B$ the EOS is difficult to distinguish numerically from the $\mu_{l,s}=0$
(i.e., $S/N_B=\infty$) case. At smaller $S/N_B$ (such as those relevant for AGS and SPS)
the energy density deviates from the zero chemical potential case at temperatures above the
crossover one.

\section{Other topics in nonzero temperature and density}

The list of subjects covered in this review is by no means exhaustive and
the chosen topics are only some of the ones closely related 
to the HIC experiments or cosmological observations.
Here I list several of the topics and recent results, which I did not cover in
my talk, and their contributors. 
\begin{itemize}
\item QGP transport coefficients:  J. Langelage
\vspace{-0.3cm}
\item Chiral magnetic effect: A. Yamamoto, T. Ishikawa
\vspace{-0.3cm}
\item  Dirac operator spectrum: F. Pittler, T. Kovacs, H. Ohno, Z. Lin
\vspace{-0.3cm}
\item $U(1)$ restoration: P. Hegde (plenary), G. Cossu 
\vspace{-0.3cm} 
\item String tension: P. Bicudo
\vspace{-0.3cm}
\item Large $N_c$ or $N_f$: M. Panero, K. Miura, M. Ogilvie, N. Yamamoto
\vspace{-0.3cm}
\item $SU(2)$ studies: P. Giudice
\vspace{-0.3cm}
\item Strong-coupling limit of QCD: O. Philipsen, S. Lottini, K. Miura,  
A. Li, etc. 
\end{itemize}    

\section{Concluding remarks}

During the last year there has been solid progress in the study of QCD at nonzero temperature and density. 
One of the important advances is the resolution of the discrepancies in the
values of $T_c$ determined with different versions of improved staggered fermions.  Another is that
more understanding was gained 
of the possible electromagnetic and finite volume effects on the $T_c$ (the latter were found in the $SU(3)$ case
only). It will be useful to extend
such calculations to the fully dynamical case and also to determine these effects on other
quantities such as the EOS.

There are some developments in the calculations of the EOS itself. 
The first one is its determination in the case of 2+1+1 dynamical flavors. The second one is
that a new method for the calculation of the EOS was introduced  and results for the
entropy density at three values of the temperature in the pure gauge case were presented.
However, in the
2+1 flavor case, the HotQCD and the WB results for the EOS at temperatures above the critical remain
quite different and currently the reason for this is unknown. The ongoing calculations with the HISQ action at 
$N_t=12$ may be helpful in the resolution of this problem.  

The mapping of the QCD phase diagram at zero and nonzero density continued this year and new
results for its properties were presented.  

There is some progress in the development of new methods for direct simulation of QCD at  
nonzero $\mu$, such as the stochastic quantization and the world-line approach; however, these methods
remain yet to be applied in the true QCD case.

Numerous studies of QCD-like theories strive to achieve at least a qualitative understanding
of the behavior of the true QCD theory in certain limits (such as the strong-coupling one) and
many new results became available this year.

In conclusion, QCD at nonzero temperature and density is a very 
dynamic area of research which encompasses a large number of subjects related
both to experimental and purely theoretical problems. This review
cannot do justice to all of them for which I apologize. I want to thank
everybody who sent me their results in advance and were available
to answer my questions. I also want to thank the organizers of this conference
for providing me with the exciting opportunity to present and discuss these topics.

\end{document}